\documentclass[runningheads,openany]{svmult}
\usepackage{palatino}
\usepackage{euler}
\usepackage{graphicx}  
\usepackage{physprbb}  
\usepackage{bg2002}  
\usepackage{epsfig}

\begin{document}

\setcounter{page}{67}

\let\I\i
\def\i{\mathrm{i}}
\def\e{\mathrm{e}}
\def\d{\mathrm{d}}
\def\half{{\textstyle{1\over2}}}
\def\thalf{{\textstyle{3\over2}}}
\def\h{{\scriptscriptstyle{1\over2}}}
\def\th{{\scriptscriptstyle{3\over2}}}
\def\fh{{\scriptscriptstyle{5\over2}}}
\def\vec#1{\mbox{\boldmath$#1$}}
\def\svec#1{\mbox{{\scriptsize \boldmath$#1$}}}
\def\oN{\overline{N}}
\def\ttimes{{\scriptstyle \times}}
\def\bm#1{{\pmb{\mbox{${#1}$}}}}

\def\CG#1#2#3#4#5#6{C^{#5#6}_{#1#2#3#4}}
\def\threej#1#2#3#4#5#6{\left(\begin{array}{ccc}
    #1&#2&#3\\#4&#5&#6\end{array}\right)}

\title*{%
Axial currents in electro-weak pion production\\
at threshold and in the $\Delta$-region
\thanks{Talk delivered by 
S. \v{S}irca.
}}

\author{%
S. \v{S}irca$^{a,b}$,
L. Amoreira$^{c,d}$,
M. Fiolhais$^{d,e}$, and
B. Golli$^{f,b}$
}\institute{%
$^a${Faculty of Mathematics and Physics,
              University of Ljubljana,
              1000 Ljubljana, Slovenia}\\
$^b${Jo\v{z}ef Stefan Institute, 
              1000 Ljubljana, Slovenia}\\
$^c${Department of Physics,
                  University of Beira Interior,
                  6201-001 Covilh\~a, Portugal}\\
$^d${Centre for Computational Physics,
                  University of Coimbra,
                  3004-516 Coimbra, Portugal}\\
$^e${Department of Physics,
                  University of Coimbra,
                  3004-516 Coimbra, Portugal}\\
$^f${Faculty of Education,
              University of Ljubljana,
              1000 Ljubljana, Slovenia}\\
}

\authorrunning{S. \v{S}irca, L. Amoreira, M. Fiolhais, and B. Golli}
\titlerunning{Axial currents in electro-weak pion production} 

\maketitle

\begin{abstract}
We discuss electro-magnetic and weak production of pions on nucleons
and show how results of experiments and their interpretation in terms
of chiral quark models with explicit meson degrees of freedom combine
to reveal the ground-state axial form factors and axial N-$\Delta$
transition amplitudes.
\end{abstract}

\section{Introduction}

The study of electro-weak N-$\Delta$ transition amplitudes,
together with an understanding of the corresponding pion
electro-production process at low energies, provides information
on the structure of the nucleon and its first excited state.
For example, the electro-magnetic transition amplitudes for the
processes $\gamma^\star\mathrm{p}\to\Delta^{+}\to\mathrm{p}\pi^0$
and $\gamma^\star\mathrm{p}\to\Delta^{+}\to\mathrm{n}\pi^+$ are
sensitive to the deviation of the nucleon shape from spherical
symmetry \cite{aron}.  Below the $\Delta$ resonance
(and in particular close to the pion-production threshold),
the reaction $\gamma^\star\mathrm{p}\to\mathrm{n}\pi^+$
also yields information on the nucleon axial and induced
pseudo-scalar form-factors.  While the electro-production of pions
at relatively high \cite{eprod_hiq} and low
\cite{eprod_loq,arnd} momentum transfers has been
intensively investigated experimentally in the past years
at modern electron accelerator facilities, very little data
exist on the corresponding weak axial processes.

\section{Nucleon axial form-factor}

In a phenomenological approach, the nucleon axial form-factor
is one of the quantities needed to extract the weak axial amplitudes
in the $\Delta$ region.  There are basically two methods to determine
this form-factor.  One set of experimental data comes from measurements
of quasi-elastic (anti)neutrino scattering on protons, deuterons,
heavier nuclei, and composite targets (see \cite{arnd} for
a comprehensive list of references).  In the quasi-elastic
picture of (anti)neutrino-nucleus scattering, the
$\nu\mathrm{N}\to\mu\mathrm{N}$ weak transition amplitude can be
expressed in terms of the nucleon electro-magnetic form-factors
and the axial form factor $G_{\mathrm{A}}$.  The axial form-factor
is extracted by fitting the $Q^2$-dependence
of the (anti)neutrino-nucleon cross section,
\begin{equation}
{{\d}\sigma\over{\d}Q^2}=
A(Q^2) \mp B(Q^2)\,(s-u) + C(Q^2)\,(s-u)^2\>{,}
\end{equation}
in which $G_{\mathrm{A}}(Q^2)$ is contained in the $A(Q^2)$, $B(Q^2)$,
and $C(Q^2)$ coefficients and is assumed to be the only unknown quantity.
It can be parameterised in terms of an `axial mass' $M_{\mathrm{A}}$ as
$$
{G_{\mathrm{A}}(Q^2) =
G_{\mathrm{A}}(0)}/(1 + Q^2/M_{\mathrm{A}}^2)^2\>{.}
$$

Another body of data comes from charged pion electro-production
on protons (see \cite{arnd} and references therein) slightly above
the pion production threshold.  As opposed to neutrino scattering,
which is described by the Cabibbo-mixed $V-A$ theory, the extraction
of the axial form factor from electro-production requires a more involved
theoretical picture \cite{BKM1,axial_review}.
The presently available most precise determination for $M_{\mathrm{A}}$
from pion electro-production is 
\begin{equation}
M_{\mathrm{A}}=(1.077\pm 0.039)\,\mathrm{GeV}
\label{MA_eprod}
\end{equation}
which is $\Delta M_{\mathrm{A}}=(0.051\pm 0.044)\,\mathrm{GeV}$ larger
than the axial mass $M_{\mathrm{A}}=(1.026\pm 0.021)\,\mathrm{GeV}$
known from neutrino scattering experiments. 
The weighted world-average estimate from electro-production data is
$M_{\mathrm{A}}=(1.069\pm 0.016)\,\mathrm{GeV}$, with an excess of
$\Delta M_{\mathrm{A}}=(0.043\pm 0.026)\,\mathrm{GeV}$ with respect
to the weak probe.  The $\sim 5\,\%$ difference in $M_{\mathrm{A}}$
can apparently be attributed to pion-loop corrections
to the electro-production process \cite{BKM1}.

\section{N-$\Delta$ weak axial amplitudes}

The experiments using neutrino scattering on deuterium or hydrogen
in the $\Delta$ region have been performed at Argonne, CERN,
and Brookhaven \cite{barish79,allen80,radecky82,kitagaki86,kitagaki90}.
(Additional experimental results exist in the quasi-elastic regime,
from which $M_{\mathrm{A}}$ has been extracted.)  For pure $\Delta$
production, the matrix element has the familiar form
$$
M = \langle\mu\Delta\,\vert\,\nu\mathrm{N}\rangle =
{G_\mathrm{F}\cos\theta_\mathrm{C}\over\sqrt{2}}\,j_\alpha\,
\langle\Delta\,\vert\,V^\alpha-A^\alpha\,\vert\,\mathrm{N}\rangle\>{,}
$$
where $G_\mathrm{F}$ is the Fermi's coupling constant, $\theta_\mathrm{C}$
is the $V_\mathrm{ud}$ element of the CKM matrix,
$j_\alpha = \overline{u}_\mu\gamma_\alpha(1-\gamma_5)u_\nu$
is the matrix element of the leptonic current, and the
matrix element of the hadronic current $J^\alpha$ has been split
into its vector and axial parts.  Typically either the $\Delta^{++}$
or the $\Delta^+$ are excited in the process.  The hadronic part
for the latter can be expanded in terms of weak vector and axial
form-factors \cite{llewellyn72}
\begin{eqnarray*}
 M = {G\over\sqrt{2}}\,
   \overline{u}_{\Delta\alpha}(p')\,\biggl\{&\,&\left[\,
   {C_3^\mathrm{V}\over M}\,\gamma_\mu
  +{C_4^\mathrm{V}\over M^2}\,p_\mu'
  +{C_5^\mathrm{V}\over M^2}\,p_\mu\,\right]\,\gamma_5 F^{\mu\alpha}
  +{C_6^\mathrm{V}}\,j_\alpha\gamma_5 \\
   + &\,& \left[\,
   {C_3^\mathrm{A}\over M}\,\gamma_\mu
  +{C_4^\mathrm{A}\over M^2}\,p_\mu'\,\right]\,F^{\mu\alpha}
  +{C_5^\mathrm{A}}\,j^\alpha
  +{C_6^\mathrm{A}\over M^2}\,q^\alpha q^\mu j_\mu\,\biggr\}\,u(p) f(W) \>{,}
\end{eqnarray*}
where $F^{\mu\alpha}=q^\mu j^\alpha - q^\alpha j^\mu$,
$\overline{u}_{\Delta\alpha}(p')$ is the Rarita-Schwinger spinor
describing the $\Delta$ state with four-vector $p'$, and $u(p)$
is the Dirac spinor for the (target) nucleon of mass $M$ with
four-vector $p$.  (In the case of the $\Delta^{++}$
excitation, the expression on the RHS acquires an additional
isospin factor of $\sqrt{3}$ since 
$\langle\Delta^{++}\,\vert\,J^\alpha\,\vert\,\mathrm{p}\rangle =
\sqrt{3}\langle\Delta^{+}\,\vert\,J^\alpha\,\vert\,\mathrm{p}\rangle =
\sqrt{3}\langle\Delta^{0}\,\vert\,J^\alpha\,\vert\,\mathrm{p}\rangle$.)
The function $f(W)$ represents a Breit-Wigner dependence
on the invariant mass $W$ of the $\mathrm{N}\pi$ system.

The matrix element is assumed to be invariant under time reversal,
hence all form-factors $C_i^\mathrm{V,A}(Q^2)$ are real.  Usually
the conserved vector current hypothesis (CVC) is also assumed to hold.
The CVC connects the matrix elements of the strangeness-conserving
hadronic weak vector current to the isovector component of
the electro-magnetic current:
\begin{eqnarray*}
\langle\Delta^{++}\,\vert\,V^\alpha\,\vert\,\mathrm{p}\rangle &=&
\sqrt{3}\,
\langle\Delta^{+}\,\vert\,J^\alpha_\mathrm{EM}(T=1)\,\vert\,
\mathrm{p}\rangle\>{,}\\
\langle\Delta^{0}\,\vert\,V^\alpha\,\vert\,\mathrm{p}\rangle &=&
\phantom{\sqrt{3}\,}
\langle\Delta^{+}\,\vert\,J^\alpha_\mathrm{EM}(T=1)\,\vert\,
\mathrm{p}\rangle\>{.}
\end{eqnarray*}
The information on the weak vector transition form-factors 
$C_i^\mathrm{V}$ is obtained from the analysis of photo-
and electro-production multipole amplitudes.  For $\Delta$
electro-excitation, the allowed multipoles are the dominant
magnetic dipole $M_{1+}$ and the electric and coulomb quadrupole
amplitudes $E_{1+}$ and $S_{1+}$, which are found to be much
smaller than $M_{1+}$ \cite{eprod_hiq,eprod_loq}.
If we assume that $M_{1+}$ dominates the electro-production
amplitude, we have $C_5^\mathrm{V} = C_6^\mathrm{V} = 0$ and
end up with only one independent vector form-factor
$$
C_4^\mathrm{V} = -{M\over W}\,C_3^\mathrm{V}\>{.}
$$
It turns out that electro-production data can be fitted well with
a dipole form for $C_3^\mathrm{V}$,
$$
C_3^\mathrm{V}(Q^2) = 2.05\,\biggl[\,
1 + {Q^2\over 0.54\,\mathrm{GeV}^2}\,\biggr]^{-2}\>{.}
$$
An alternative parameterisation of $C_3^\mathrm{V}$ which accounts
for a small observed deviation from the pure dipole form is
$$
C_3^\mathrm{V}(Q^2) = 2.05\,\Bigl[\, 1 + 9 \sqrt{Q^2} \,\Bigr]\,
\exp\Bigl[\,-6.3 \sqrt{Q^2}\,\Bigr]\>{.}
$$
The main interest therefore lies in the axial part of the hadronic
weak current which is not well known.

\subsection*{Extraction of $C_i^\mathrm{A}(Q^2)$ from data}

The key assumption in experimental analyses of the axial matrix
element is the PCAC.  It implies that the divergence of the axial
current should vanish as $m_\pi^2\to 0$, which occurs
if the induced pseudo-scalar term with $C_6^\mathrm{A}$
(the analogue of $G_\mathrm{P}$ in the nucleon case) is dominated
by the pion pole.  In consequence, $C_6^\mathrm{A}$ can be expressed
in terms of the strong $\pi\mathrm{N}\Delta$ form-factor,
$$
{C_6^\mathrm{A}(Q^2)\over M^2} = f_\pi\sqrt{2\over 3}\,
{G_{\pi\mathrm{N}\Delta}\over 2M}\,{1\over Q^2 + m_\pi^2}\>{,}
$$
while $C_5^\mathrm{A}$ and $C_6^\mathrm{A}$ can be approximately
connected through the off-diagonal Gold\-ber\-ger-Treiman relation
\cite{miniBled}.  In a phenomenological analysis, $C_3^\mathrm{A}(Q^2)$,
$C_4^\mathrm{A}(Q^2)$, and $C_5^\mathrm{A}(Q^2)$ are taken as free
parameters and are fitted to the data.  The axial form-factors are
also parameterised in ``corrected'' dipole forms
$$
C_i^\mathrm{A}(Q^2) = C_i^\mathrm{A}(0)\,
\biggl[\,1 + {a_i Q^2\over b_i + Q^2}\,\biggr]\,
\biggl[\,1 + {Q^2\over M_\mathrm{A}^2}\,\biggr]^{-2}\>{.}
$$
In the simplest approach one takes $a_i=b_i=0$.  Historically,
the experimental data on weak pion production could be understood
well enough in terms of a theory developed by Adler \cite{adler}.
For lack of a better choice, Adler's values for $C_i^\mathrm{A}(0)$
have conventionally been adopted to fix the fit-parameters at $Q^2=0$,
i.~e.
\begin{eqnarray}
C_3^\mathrm{A}(0) &=&  0\>{,}\\
C_4^\mathrm{A}(0) &=& -0.3\>{,}\label{C4A_adler}\\
C_5^\mathrm{A}(0) &=&  1.2\>{.}
\end{eqnarray}
In such a situation, one ends up with $M_\mathrm{A}$ as
the only free fit-parameter.

Several observables are used to fit the $Q^2$-dependence
of the form-factors.  Most commonly used are the total cross-sections
$\sigma(E_\nu)$, and the angular distributions of the recoiling
nucleon
$$
{\mathrm{d}\sigma\over\mathrm{d}\Omega} = {\sigma\over\sqrt{4\pi}}\,\biggl[\,
Y_{00} 
 - {2\over\sqrt{5}}\Bigl[\,\tilde{\rho}_{33}-{1\over 2}\,\Bigr]\,Y_{20}
 + {4\over\sqrt{10}}\,\tilde{\rho}_{31}\,\mathrm{Re}Y_{21}
 - {4\over\sqrt{10}}\,\tilde{\rho}_{3-1}\,\mathrm{Re}Y_{22}\,\biggr]\>{,}
$$
where $\tilde{\rho}_{mn}$ are the density matrix elements and 
$Y_{LM}$ are the spherical harmonics.   Better than from the
$\tilde{\rho}_{mn}$ coefficients, the $Q^2$ dependence of the
matrix element can be determined from the differential cross-section
$\mathrm{d}\sigma/\mathrm{d}Q^2$.  In particular, since the
dependence on $C_3^\mathrm{A}$ and $C_4^\mathrm{A}$ is anticipated
to be weak at $Q^2\sim 0$, then 
$$
{\mathrm{d}\sigma\over\mathrm{d}Q^2}(Q^2=0) 
\propto (\,C_5^\mathrm{A}(0)\,)^2\>{.}
$$

The refinements of this crude approach are dictated by several
observations.  If the target is a nucleus (for example, the
deuteron which is needed to access specific charge channels),
nuclear effects need to be estimated.  Another important
correction arises due to the finite energy width of the $\Delta$.
In addition, the non-zero mass of the scattered muon may play
a role at low $Q^2$.

All these effects have been addressed carefully in \cite{valencia}.
The sensitivity of the differential cross-section to different
nucleon-nucleon potentials was seen to be smaller than $10\,\%$
even at $Q^2<0.1\,\mathrm{GeV}^2$.  In the range above that value,
this allows one to interpret inelastic data on the deuteron 
as if they were data obtained on the free nucleon.  The effect
of non-zero muon mass is even less pronounced: it does not
exceed $5\,\%$ in the region of $Q^2\sim 0.05\,\mathrm{GeV}^2$.
The energy dependence of the width of the $\Delta$ resonance
was observed to have a negligible effect on the cross-section.
The final value based on the analysis of Argonne data
\cite{radecky82} is
\begin{equation}
C_5^\mathrm{A}(0) = 1.22 \pm 0.06\>{.}
\label{C5A0_uncert}
\end{equation}

\begin{figure}[h]
\begin{center}
\includegraphics[width=60mm]{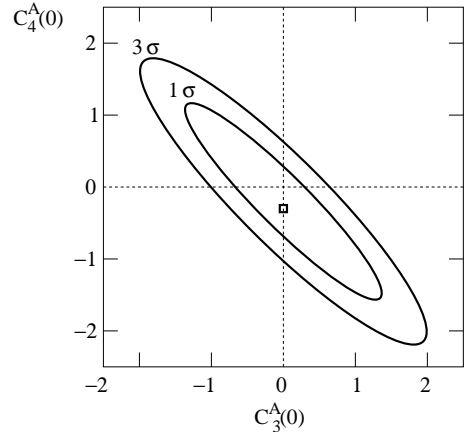}
\end{center}
\caption{One- and three-standard deviation limits on $C_3^\mathrm{A}(0)$
and $C_4^\mathrm{A}(0)$ as extracted from measurements of
$\nu_\mu\mathrm{p}\to\mu^-\Delta^{++}$.  The square denotes
the model predictions by Adler \protect\cite{adler}.
(Figure adapted after \protect\cite{barish79}.)}
\label{barish79_fig16}
\end{figure}

At present, this is the best estimate for $C_5^\mathrm{A}(0)$,
although a number of phenomenological predictions also exist
\cite{mukh98}.  We adopt this value for the purpose of
comparison to our calculations.  There is also some scarce,
but direct experimental evidence from a free fit to the data
that $C_3^\mathrm{A}(0)$ is indeed small and $C_4^\mathrm{A}(0)$
is close to the Adler's value of $-0.3$ (see Figure~\ref{barish79_fig16}).
We use $C_4^\mathrm{A}(0)=-0.3$ in our comparisons in the next section.

\section{Interpretation of $C_i^\mathrm{A}(Q^2)$ in the linear
$\sigma$-model}

The axial N-$\Delta$ transition amplitudes can be interpreted
in an illustrative way in quark models involving chiral fields
like the linear $\sigma$-model (LSM), which may reveal the importance
of non-quark degrees of freedom in baryons.
Due to difficulties in consistent incorporation
of the pion field, the model predictions for these amplitudes
are very scarce \cite{Mukh}.  The present work \cite{FB18,letter02}
was partly also motivated by the experience gained in the successful
phenomenological description of the quadrupole
electro-excitation of the $\Delta$ within the LSM,
in which the pion cloud was shown to play a major role \cite{FGS}.

\subsection{Two-radial mode approach}

We have realised that by treating the nucleon and the $\Delta$
in the LSM in a simpler, one-radial mode ansatz, the off-diagonal
Goldberger-Treiman relation can not be satisfied.  For the calculation
of the amplitudes in the LSM, we have therefore used the two-radial
mode ansatz for the physical baryon states which allows for
different pion clouds around the bare baryons.  The physical
baryons are obtained from the superposition of bare quark cores
and coherent states of mesons by the Peierls-Yoccoz angular
projection.  For the nucleon we have the ansatz
\begin{equation}
  |\mathrm{N}\rangle = \mathcal{N}_\mathrm{N}
  P^\h\left[\Phi_\mathrm{N}|\mathrm{N}_q\rangle +
  \Phi_{\mathrm{N}\Delta}|\Delta_q\rangle\right] \>{,}
\label{N2mode}
\end{equation}
where $\mathcal{N}_\mathrm{N}$ is the normalisation factor.
Here $\Phi_\mathrm{N}$ and $\Phi_{\mathrm{N}\Delta}$ stand for hedgehog
coherent states describing the pion cloud around the bare nucleon
and bare $\Delta$, respectively, and $P^\h$ is the projection operator
on the subspace with isospin and angular momentum $\half$.
Only one profile for the $\sigma$ field is assumed.
For the $\Delta$ we assume a slightly different ansatz 
to ensure the proper asymptotic behaviour.  We take
\begin{equation}
  |\Delta\rangle = \mathcal{N}_\Delta\left\{
   P^\th\Phi_\Delta|\Delta_q\rangle +
   \int\d k\, \eta(k)
[a_{mt}^\dagger(k)|\mathrm{N}\rangle
]^{\th\th}\right\}\,,
\label{D2mode}
\end{equation}
where $\mathcal{N}_\Delta$ is the normalisation factor,
$|\mathrm{N}\rangle$ is the ground state and
$[\,\,\,]^{\th\th}$ denotes the pion-nucleon state with
isospin $\thalf$ and spin $\thalf$. We have
interpreted the localised model states as wave-packets
with definite linear momentum, as elaborated in \cite{miniBled}.

\subsection{Calculation of helicity amplitudes}

We use the kinematics and notation of \cite{miniBled}.  For the quark
contribution to the two transverse ($\lambda=1$) and longitudinal
($\lambda=0$) helicity amplitudes we obtain
\begin{eqnarray*}
\tilde{A}^{\mathrm{(q)}}_{s_\Delta\lambda} &=& -\langle\Delta_{s_\Delta\,\h}|
  \int\d\vec{r}\, e^{\i kz}\psi^\dagger\alpha_\lambda\gamma_5\half\tau_0\psi
|\mathrm{N}_{s_\Delta-\lambda\,\h}\rangle\\
 \tilde{A}^{(q)}_{s_\Delta\lambda} 
  &=& 
    - \half\mathcal{N}_\Delta \int\d r\,r^2\biggl\{\\
  &\,&\biggl[j_0(kr)
      \biggl(u_\Delta u_\mathrm{N} - {1\over3} v_\Delta v_\mathrm{N}\biggr)
    +{2\over3}\,(3\lambda^2-2)\,j_2(kr)v_\Delta v_\mathrm{N}\biggr]
	\langle\Delta_b||\sigma\,\tau||\mathrm{N}\rangle \\
  - &c_\eta&   \biggl[j_0(kr)
      \biggl(u_\mathrm{N}^2 - {1\over3} v_\mathrm{N}^2\biggr)
    +{2\over3}\,(3\lambda^2-2)\,j_2(kr)v_\mathrm{N}^2\biggr]
\nonumber\\
  &&\times
\biggl[{4\over9} \langle \mathrm{N}||\sigma\tau||\mathrm{N}\rangle
             + {1\over36}\langle \mathrm{N}||\sigma\tau||\mathrm{N}(J=\thalf)\rangle
               \biggr]
	\biggr\}
             \CG{\half}{s_\Delta-\lambda}{1}{\lambda}{\thalf}{s_\Delta}
             \CG{\half}{\half}{1}{0}{\thalf}{\half} \>{.}
\end{eqnarray*}
Here $u$ and $v$ are upper and lower components of Dirac spinors for
the nucleon and the $\Delta$, while $c_\eta$ is a coefficient
involving integrals of the function $\eta(k)$ appearing in
(\ref{D2mode}).  The reduced matrix elements of $\vec{\sigma}\vec{\tau}$
can be expressed in terms of analytic functions with intrinsic numbers
of pions as arguments.  In all three cases, we take $s_\Delta=\thalf$.
For the scalar amplitudes, we take $\lambda=0$ and $s_\Delta=\half$,
and obtain
\begin{eqnarray*}
\tilde{S}^{\mathrm{(q)}} &=& - \langle\Delta_{\h\h}|
  \int\d\vec{r}\, e^{\i kz}\psi^\dagger\gamma_5\half\tau_0\psi
|\mathrm{N}_{\h\h}\rangle
\label{Sqdef}\\
 &=&  {1\over3}\, \mathcal{N}_\Delta \int\d r\,r^2\,j_1(kr)
      \left(u_\Delta v_\mathrm{N} - v_\Delta u_\mathrm{N}\right)
       \langle\Delta_b||\sigma\tau||\mathrm{N}\rangle\;.
\label{Sq}
\end{eqnarray*}

For the non-pole meson contribution to the transverse and longitudinal
helicity amplitudes we assume the same $\sigma$ profiles
around the bare states, but different for the physical states.
Introducing an ``average'' $\sigma$ field 
$\bar{\sigma}(r)\equiv\half(\sigma_\mathrm{N}(r) + \sigma_\Delta(r))$
we obtain
\begin{eqnarray*}
\tilde{A}^{\mathrm{(m)}}_{s_\Delta\lambda} &=& \langle\Delta_{s_\Delta\,\h}|
  \int\d\vec{r}\, e^{\i kz}
  \left((\sigma-f_\pi)\nabla_\lambda\pi_0-\pi_0\nabla_\lambda\sigma\right)
  |\mathrm{N}_{s_\Delta-\lambda\,\h}\rangle\\
  &=& 
  {4\pi\over3}
     \Biggl\{\int\d r\,r^2\, j_0(kr)\Biggl[\Biggl((\bar{\sigma}-f_\pi)
     \left({\d\varphi_{\Delta \mathrm{N}}\over\d r} 
             + {2\varphi_{\Delta \mathrm{N}}\over r}\right)
     - {\d\bar{\sigma}\over\d r}\,\varphi_{\Delta \mathrm{N}}\Biggr)
      \Biggr]\Biggr.
\nonumber\\
  && \Biggl. \hspace{-2mm}
  + (3\lambda^2-2)\int\d r\,r^2\, j_2(kr) \Biggl[\Biggl((\bar{\sigma}-f_\pi)
    \left({\d\varphi_{\Delta \mathrm{N}}\over\d r} - {\varphi_{\Delta \mathrm{N}}\over r}\right)
     - {\d\bar{\sigma}\over\d r}\,\varphi_{\Delta \mathrm{N}}\Biggr)
      \Biggr]\Biggr\}  
\nonumber\\
  && \times \CG{\half}{s_\Delta-\lambda}{1}{\lambda}{\thalf}{s_\Delta}
             \CG{\half}{\half}{1}{0}{\thalf}{\half}\>{,}
\label{Am}
\end{eqnarray*}
where
$\varphi_{\Delta\mathrm{N}}=\langle\Delta\,|\,\pi\,|\,\mathrm{N}\rangle$.
To compute the scalar amplitude,
we make use of the off-diagonal virial relation derived in
\cite{miniBled} and define
$$
 \sigma^P(r) 
      = \int_0^\infty \d k\, k^2\sqrt{k^2+m_\sigma^2} \sqrt{2\over\pi} 
   \,j_0(kr)\, \sigma(k)\>{.}
$$
We obtain
\begin{eqnarray*}
\tilde{S}^{\mathrm{(m)}} &=&
 - \langle\Delta_{\h\h}|
  \int\d\vec{r}\, e^{\i kz}
  \left((\sigma-f_\pi) P_{\pi0}-P_\sigma\pi_0\right)
  |\mathrm{N}_{\h\h}\rangle\\
 &=&  - {8\pi\over3}\int\d r\, r^2\,j_1(kr)\left\{
     \half\left(\sigma^P_\mathrm{N}(r) - \sigma^P_\Delta(r)\right)\varphi_{\Delta \mathrm{N}}(r)
  - \left(\bar{\sigma}(r)-f_\pi\right)\omega_*\varphi_{\Delta \mathrm{N}}(r) 
  \right\} \>{.}
\end{eqnarray*}
By using
$$
\tilde{A}_{s_\Delta\lambda} 
           = \left(A^0 - (3\lambda^2-2)A^2\right)
             \CG{\half}{s_\Delta-\lambda}{1}{\lambda}{\thalf}{s_\Delta}
             \CG{\half}{\half}{1}{0}{\thalf}{\half}\>{,}
$$
the quark and non-pole meson contributions to the transverse
amplitudes can finally be broken into $L=0$ and $L=2$ pieces,
\begin{eqnarray}
\tilde{A}^\mathrm{A}_\th &=& \sqrt{2\over3}\left(A^0 - A^2\right)\>{,}
\label{A32}\\
\tilde{A}^\mathrm{A}_\h &=& {1\over\sqrt{3}} \tilde{A}^\mathrm{A}_\th =
                 {\sqrt{2}\over3}\left(A^0 - A^2\right)\>{,}
\label{A12}\\
 \tilde{L}^\mathrm{A} &=& {2\over3}\left(A^0 + 2 A^2\right) \>{,}
\label{L}
\end{eqnarray}
and inserted into (76), (77), and (78) of \cite{miniBled}.
The pole part of the meson contribution is
$$
 C^\mathrm{A}_{6\,\mathrm{(pole)}}(Q^2)
= f_\pi {G_{\pi\mathrm{N}\Delta}(Q^2)\over 2M_\mathrm{N}}\,
       {M_\mathrm{N}^2\over m_\pi^2 + Q^2}\,\sqrt{{2\over3}} \>{.}
$$
The strong $\mathrm{N}\Delta$ form-factor $G_{\pi\mathrm{N}\Delta}$
can be computed through
$$
{G_{\pi\mathrm{N}\Delta}(Q^2)\over 2M_\mathrm{N}}
{M_\Delta + M_\mathrm{N}\over 2M_\Delta} =
{1\over\mathrm{i}\,k}\,\langle\Delta\parallel\,
\int\mathrm{d}\vec{r}\, \mathrm{e}^{\mathrm{i}\svec{k}\svec{r}}\,J(\vec{r})\,
\parallel\mathrm{N}\rangle\>{,}
$$
where the current $J$ has a component corresponding to the quark source
and a component originating in the meson self-interaction term
(see (58) of \cite{miniBled}),
$$
  J_0(\vec{r}) = j_0(\vec{r}) +
      {\partial U(\sigma,\pol{\pi})\over\partial\pi_0(\vec{r})}\>{.}
$$

\subsection{Results}

\begin{figure}[h]
\begin{center}
\includegraphics[height=80mm]{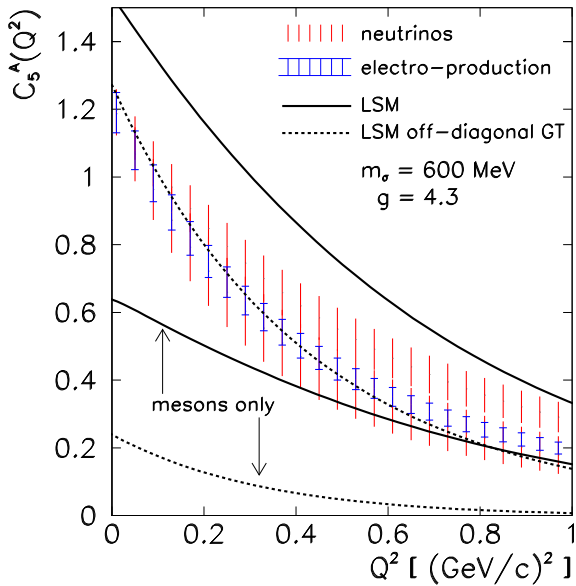}
\end{center}
\caption{The amplitude $C_5^\mathrm{A}(Q^2)$ in the two-radial
mode LSM.  The experimental uncertainty at $Q^2=0$ is given by
Eq.~(\protect\ref{C5A0_uncert}).  The error ranges are given
by the spread in the axial-mass parameter $M_\mathrm{A}$
as determined from neutrino scattering experiments 
(broader range, \cite{kitagaki90}) and from electro-production
of pions (narrower range, Eq.~(\protect\ref{MA_eprod})).
Full curves: calculation from helicity amplitudes
(\protect\ref{A32}), (\protect\ref{A12}), and (\protect\ref{L});
dashed curves: calculation from $G_{\pi\mathrm{N}\Delta}$.}
\label{fig:c5a}
\end{figure}

Fig.~\ref{fig:c5a} shows the $C^\mathrm{A}_5(Q^2)$ amplitude 
with the quark-meson coupling constant of $g=4.3$ and
$m_\sigma=600\,\mathrm{MeV}$ compared to the experimentally
determined form-factors.  The figure also shows the $C^\mathrm{A}_5(Q^2)$
calculated from the strong $\pi\mathrm{N}\Delta$ form-factor
using the off-diagonal Goldberger-Treiman relation.

The magnitude of $C_5^\mathrm{A}(Q^2)$ is overestimated
in the LSM, with $C_5^\mathrm{A}(0)$ about $25\,\%$ higher than
the experimental average.  Still, the $Q^2$-dependence
follows the experimental one very well: the $M_\mathrm{A}$
from a dipole fit to our calculated values agrees to within
a few percent with the experimental $M_\mathrm{A}$.
On the other hand, with $C^\mathrm{A}_5(Q^2)$ determined
from the calculated strong $\pi\mathrm{N}\Delta$ form-factor,
the absolute normalisation improves, while the $Q^2$ fall-off 
is steeper, with $M_\mathrm{A}\approx 0.80\,\mathrm{GeV}$.
Since the model states are not exact eigenstates of the LSM Hamiltonian,
the discrepancy between the two calculated values in some sense
indicates the quality of the computational scheme.  At $Q^2=-m_\pi^2$
where the off-diagonal Goldberger-Treiman relation is expected to hold,
the discrepancy is $17\,\%$.  The disagreement between the two
approaches can be attributed to an over-estimate of the meson
strength, a characteristic feature of LSM where only the meson
fields bind the quarks.

Essentially the same trend is observed in the ``diagonal'' case:
for the nucleon we obtain $g_\mathrm{A}=1.41$.  The discrepancy
with respect to the experimental value of $1.27$ is commensurate
with the disagreement in $C_5^\mathrm{A}(0)$.
The overestimate of $g_\mathrm{A}$ and $G_\mathrm{A}(Q^2)$
was shown to persist even if the spurious centre-of-mass motion
of the nucleon is removed \cite{fizika383}.  An additional
projection onto non-zero linear momentum therefore does not
appear to be feasible.

The effect of the meson self-interaction is relatively less
pronounced in the strong coupling constant
(only $\sim 20\,\%$) than in $C^\mathrm{A}_5(Q^2)$.
Both $G_{\pi\mathrm{N}\Delta}(0)$ and $G_{\pi\mathrm{NN}}(0)$
are over-estimated in the model by $\sim 10\,\%$.  Still,
the ratio $G_{\pi\mathrm{N}\Delta}(0)/G_{\pi\mathrm{NN}}(0)=2.01$
is considerably higher than either the familiar SU(6)
prediction $\sqrt{72/25}$ or the mass-corrected value
of $1.65$~\cite{Hemmert}, and compares reasonably well
with the experimental value of $2.2$.  This improvement
is mostly a consequence of the renormalisation of the strong
vertices due to pions.

\begin{figure}[h]
\begin{center}
\includegraphics[height=80mm]{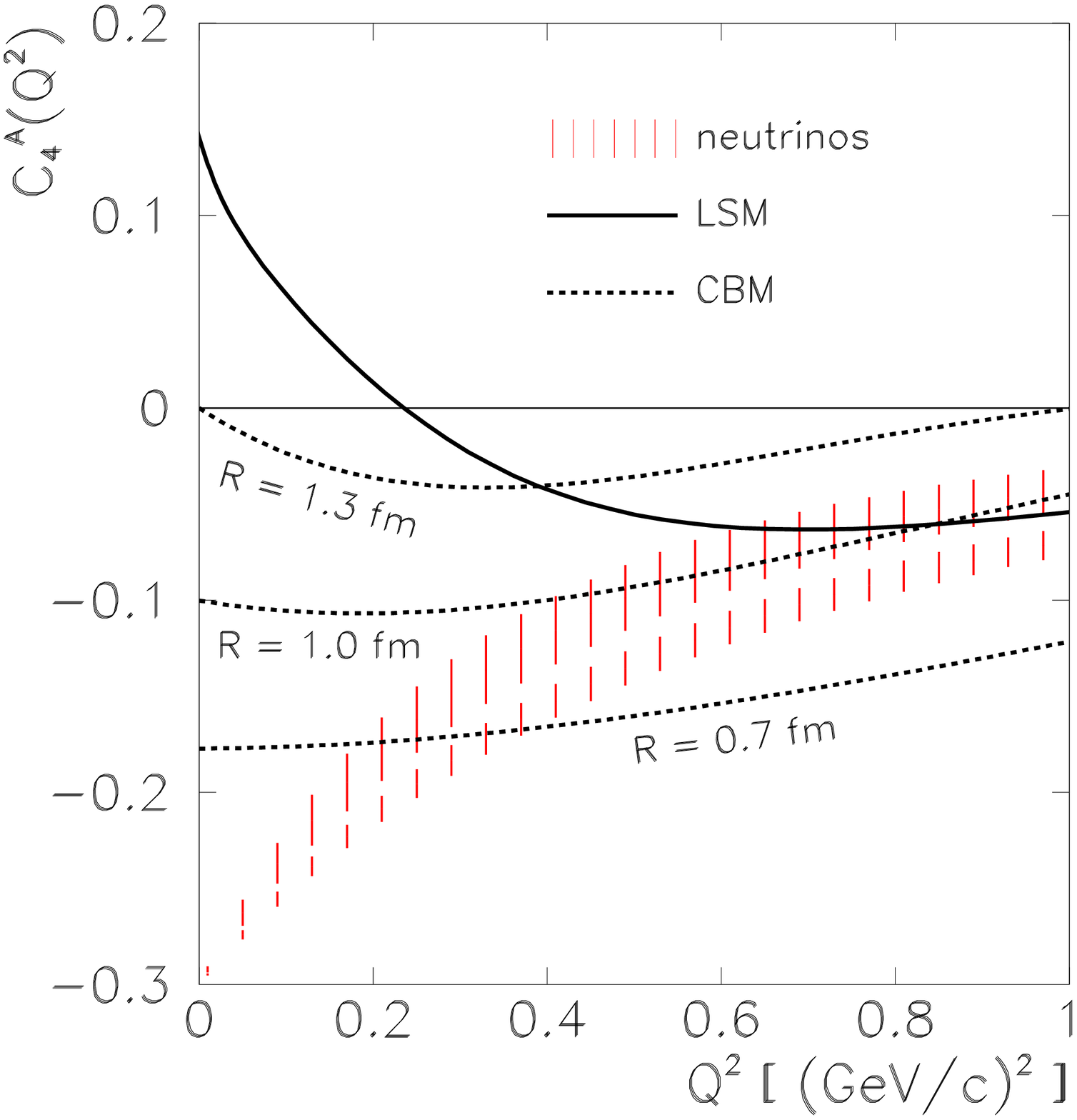}
\end{center}
\vspace*{-2mm}
\caption{The amplitude $C_4^\mathrm{A}(Q^2)$ in the two-radial
mode linear $\sigma$-model, with model parameters and experimental
uncertainties due to the spread in $M_\mathrm{A}$ as in
Fig.~\protect\ref{fig:c5a}, and in the Cloudy-Bag Model
(see below for discussion).  Experimentally,
$C^\mathrm{A}_4(0)=-0.3\pm 0.5$ (see \protect\cite{barish79}
and Fig.~\protect\ref{barish79_fig16}).  For orientation,
the value for $C_4^\mathrm{A}(0)$ is used without error-bars.}
\label{fig:c4a}
\end{figure}

The determination of the $C^\mathrm{A}_4(Q^2)$ is less reliable
because the meson contribution to the scalar component
of this amplitude \cite{miniBled} is very sensitive
to small variations of the profiles.
However, the experimental value is very uncertain as well.
Neglecting the non-pole contribution to the scalar amplitude
and $C_6^\mathrm{A}(Q^2)$ (with the pole contribution canceling out),
$C_4^\mathrm{A}(Q^2)$ is fixed to
$-(M_\mathrm{N}^2/2M_\Delta^2)\,C_5^\mathrm{A}(Q^2)$.
At $Q^2=0$, this is in excellent numerical agreement with (\ref{C4A_adler}).
In the LSM, the non-pole contribution to $C_6^\mathrm{A}(Q^2)$
happens to be non-negligible and tends to increase $C_4^\mathrm{A}(Q^2)$
at small $Q^2$, as seen in Fig.~\ref{fig:c4a}.  An almost identical
conclusion regarding $C^\mathrm{A}_4(Q^2)$ applies in the case
of the Cloudy-Bag Model, as shown below.

\begin{figure}[hb]
\begin{center}
\includegraphics[height=70mm]{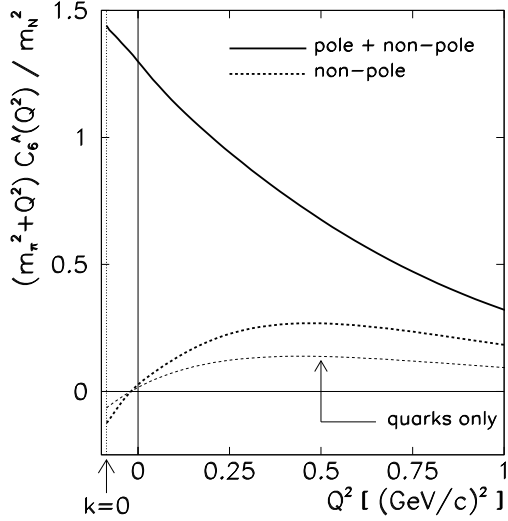}
\end{center}
\vspace*{-2mm}
\caption{The non-pole part and the total amplitude
$C_6^\mathrm{A}(Q^2)$ in the two-radial mode linear $\sigma$-model.
Model parameters are as in Fig.~\protect\ref{fig:c5a}.}
\label{fig:c6a}
\end{figure}

The $C^\mathrm{A}_6$ amplitude is governed by the pion pole
for small values of $Q^2$ and hence by the value
of $G_{\pi\mathrm{N}\Delta}$ which is well reproduced
in the LSM, and underestimated by $\sim 35\,\%$ in the Cloudy-Bag
Model.  Fig.~\ref{fig:c6a} shows that the non-pole contribution
becomes relatively more important at larger values of $Q^2$.

\section{Interpretation of $C_i^\mathrm{A}(Q^2)$ in the Cloudy-Bag Model}

For the calculation in the Cloudy-Bag Model (CBM) we have assumed
the usual perturbative form for the pion profiles using the experimental
masses for the nucleon and $\Delta$.  Since the pion contribution
to the axial current in the CBM has the form
$f_\pi\partial^\alpha\vec{\pi}$, only the quarks contribute
to the $C^\mathrm{A}_4(Q^2)$ and $C^\mathrm{A}_5(Q^2)$,
while $C^\mathrm{A}_6(Q^2)$ is almost completely dominated
by the pion pole (see contribution by B.~Golli \cite{miniBled}).
With respect to the LSM, the sensitivity
of the axial form-factors to the non-quark degrees of freedom
is therefore almost reversed.

\begin{figure}[h]
\begin{center}
\includegraphics[height=80mm]{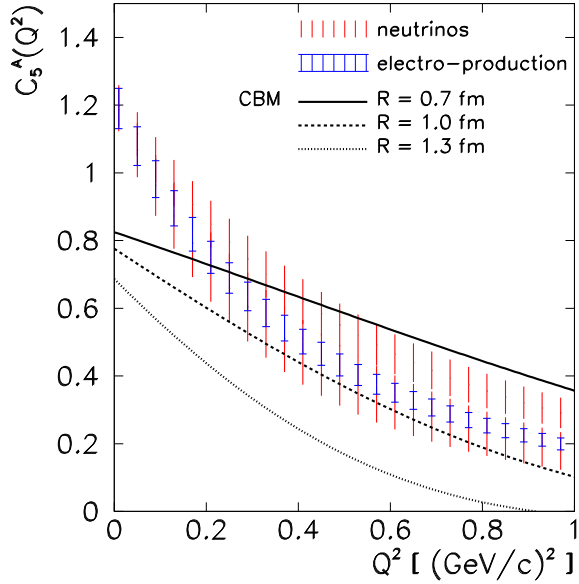}
\end{center}
\caption{The amplitude $C_5^\mathrm{A}(Q^2)$ in the Cloudy-Bag Model
for three values of the bag radius.  Experimental uncertainties
are as in caption to Fig.~\protect\ref{fig:c5a}.}
\label{fig:c5a_CBM}
\end{figure}

In the CBM, only the non-pole
component of the axial current contributes to the amplitudes,
and as a result the $C^\mathrm{A}_5(0)$ amplitude is less than $2/3$
of the experimental value.  The behaviour of $C^\mathrm{A}_5(Q^2)$
(see Fig.~\ref{fig:c5a_CBM}) is similar as in the pure MIT Bag Model
(to within $10\,\%$),
with fitted $M_\mathrm{A}\sim 1.2\,\mathrm{GeV\,fm}/R$.
The off-diagonal Goldberger-Treiman relation is satisfied
in the CBM, but $C^\mathrm{A}_5$ from $G_{\pi\mathrm{N}\Delta}$
has a steeper fall-off with fitted
$M_\mathrm{A}\sim 0.8\,\mathrm{GeV\,fm}/R$.

The large discrepancy can be partly attributed to the
fact that the CBM predicts a too low value for 
$G_{\pi\mathrm{NN}}$, and consequently $G_{\pi\mathrm{N}\Delta}$.
We have found that the pions
increase the $G_{\pi\mathrm{N}\Delta}/G_{\pi\mathrm{NN}}$ ratio
by $\sim 15\,\%$ through vertex renormalisation.
The effect is further enhanced by the mass-correction factor
$2M_\Delta/(M_\Delta+M_\mathrm{N})$, yet suppressed in
the kinematical extrapolation of $G_{\pi\mathrm{N}\Delta}(Q^2)$
to the $\mathrm{SU}(6)$ limit.  This suppression is weaker
at small bag radii $R$: the ratio drops from $2.05$
at $R=0.7\,\mathrm{fm}$ to $1.60$ (below the $\mathrm{SU}(6)$
value) at $R=1.3\,\mathrm{fm}$.

The determination of the $C^\mathrm{A}_4(Q^2)$ is less reliable
for very much the same reason as in the LSM.  The non-pole contribution
to $C_6^\mathrm{A}(Q^2)$ tends to add to the excessive strength of
$C_4^\mathrm{A}(Q^2)$ at low $Q^2$, as seen in Fig.~\ref{fig:c4a}.
Never the less, the experimental data are too coarse to allow
for a meaningful comparison to the model.  For technical details
regarding the calculation in the CBM, refer to \cite{miniBled}.


\end{document}